\documentclass[letter,twocolumn,prl,showpacs]{revtex4}

\usepackage[dvips]{graphicx} 
\usepackage{amssymb} 
\usepackage{amsmath} 
\usepackage{latexsym} 

\newcommand\be{\begin{equation}}
\newcommand\bea{\begin{eqnarray}}
\newcommand\eea{\end{eqnarray}}
\newcommand\ee{\end{equation}}

\begin{document}


\title{Spontaneous Patterning of Confined Granular Rods}

\author{Jennifer Galanis$^1$, Daniel Harries$^1$, Dan L. Sackett$^1$, 
Wolfgang Losert$^2$ and Ralph Nossal$^1$}

\affiliation{$^1$NICHD, National Institutes of Health, Bethesda, Maryland 20892-0924, USA}

\affiliation{$^2$Department of Physics, IPST, and IREAP, University of Maryland, College Park, Maryland 20742, USA}


\begin{abstract}

Vertically vibrated rod-shaped granular materials confined to quasi-2D containers self organize into distinct patterns. We find, consistent with theory and simulation, a density dependent isotropic-nematic transition. Along the walls, rods interact sterically to form a  wetting layer. For high rod densities, complex patterns emerge as a result of competition between bulk and boundary alignment. A continuum elastic energy accounting for nematic distortion and local wall anchoring reproduces the structures seen experimentally.

\end{abstract}

\maketitle


From nanometer sized molecules forming liquid crystals to timber floating down a river, rod shaped materials pervade our everyday existence. Onsager first demonstrated that hard core steric interactions suffice to order rods in thermal equilibrium \cite{onsager}. Above a critical concentration, orientational entropy is sacrificed for gains in translational entropy; therefore, crowding alone induces a temperature-independent isotropic-nematic (I-N) transition \cite{LC}. In this entropically driven transition, the kinetic motion of the rods serves as a mechanism for adequately sampling phase space. While thermal excitation mixes microscopic particles, macroscopic rods require an externally applied energy for randomization. For both vertical and horizontal vibrations, a density dependent nematic to smectic-like transition was previously observed for granular rods \cite{kudrolli03}. Such driven dissipative systems, however, are far from equilibrium and need not evolve to states of maximal configurational entropy. For example, prior studies utilizing macroscopic rods focused on jammed or other metastable states \cite{kudrolli03, granular}.

Here, we report on finite-sized driven dissipative granular systems of rods in steady-state that share many properties of lyotropic liquid crystals at equilibrium. We find that short ranged hard core interactions determine the arrangement of the moving rods for all densities. A phase transition from a disordered isotropic state to a nematic-like state occurs as a function of rod density and rod aspect (length-to-diameter) ratio, $L/D$. By using excluded volume scaling predicted by mean field theories for liquid crystals, our data give a single normalized transition density for all rod $L/D$ and densities tested.

However, excluded volume interactions in the bulk cannot explain all rod behavior because confining boundaries also strongly influence rod ordering. For relatively low rod densities, experimental results for rod alignment and density profiles with respect to the wall strikingly resemble theoretical predictions \cite{poniewierski}, indicating that rods interact via hard core steric repulsions with the surrounding boundaries. As the rods become dense, rod ordering in the bulk nematic competes with rod ordering along the surrounding walls. This frustration creates distinct yet easily reproduced patterns. We account for observed structures by modeling the system as a liquid crystal undergoing the elastic deformations of splay and bend subject to a simple wall interaction \cite{frank,deGennes}. Moreover, we show that the patterns depend on the relative magnitude of wall and bulk energies.


Rod-shaped granular materials confined to quasi-$2$D containers were vertically vibrated. Stainless steel wire, $0.08\,\rm{cm}$ diameter, was cut into rods of $L/D=20$, $40$, and $60$. Containers were either circular (radius $R=15\,\rm{cm}$ or $R=7\,\rm{cm}$) or square ($29\,\rm{cm}$ diagonal). The chamber height to rod diameter ratio was $20$, which restricted the rods to lie mainly flat. Due to the quasi-$2$D nature of these experiments, rod densities were scaled to the area of the container bottom, $A$, so that area fraction is $\phi=NLD/A$, where $N$ is the number of rods. A monolayer of rods lying flat and fully covering the container bottom corresponds to $\phi=1$.

The rods were shaken with a sinusoidal acceleration of $50\,\rm{Hz}$ and a peak acceleration of $4$ times gravity. The magnetic field of the shaker was minimized by distancing the rod chamber from the shaker. Electrostatic forces were reduced by using aluminum chamber components and applying an anti-static coating to non-metallic surfaces. The apparatus was carefully leveled. To randomize initial conditions, rods were dropped into the containers from above before experiments commenced. Images of the rods in motion were captured by a high speed digital camera. The position and orientation of each rod was found by an image processing algorithm, developed in Matlab $6.5$. The counting accuracy was close to $100\%$ for low density systems and greater than $75\%$ for samples as dense as $\phi = 0.6$.


Unlike the infinite bulk modeled by Onsager's mean field theory \cite{onsager}, rigid boundaries confine our experimental system. We first investigate the I-N transition away from these strong wall influences. By measuring the angular correlation of the rods with respect to the nearest wall [wall-rod correlation, $g_{2}(r)$], we distinguish rods affected by the side boundaries from those that were not. Briefly, we define an effective ``bulk'' when $g_{2}(r)\approx 0$, which occurs at distances $r/L>0.5$ from the wall. We will later discuss in detail $g_{2}(r)$ and other parameters in relation to ordering along the boundaries.

To measure rod ordering in the bulk, we characterize alignment via the order parameter $S= \langle\,\cos(2\theta_{i})\,\rangle $, where $\theta_{i}$ is the angle of the $i$th rod with the nematic director. In practice, the order parameter and director are found for an arbitrary Cartesian coordinate system via the matrix $Q= \langle\, 2 u_{\alpha}(i) u_{\beta}(i) - \delta_{\alpha\beta} \,\rangle $, where $u_{\alpha}(i)$ and $u_{\beta}(i)$ are the coordinates of the unit vector specifying the direction of the $i$th rod. The eigenvalues of $Q$ give $\pm S$, and the eigenvectors give the director and its perpendicular vector. A perfect nematic corresponds to $S=1$. Standard convention denotes an isotropic system as $S=0$, but by the way $S$ is defined, an isotropic system will have $S>0$ that approaches $0$ only as $N \to \infty$ \cite{frenkel85}. To distinguish rod ordering from this finite-number effect, we evaluated $S$ of finite numbers of non-interacting rods using 2D Monte Carlo (MC) simulations, where the number of rods in the simulations matched those in experiments.

\begin{figure}[!bp]
  \includegraphics[keepaspectratio=true,width=0.29\textwidth]{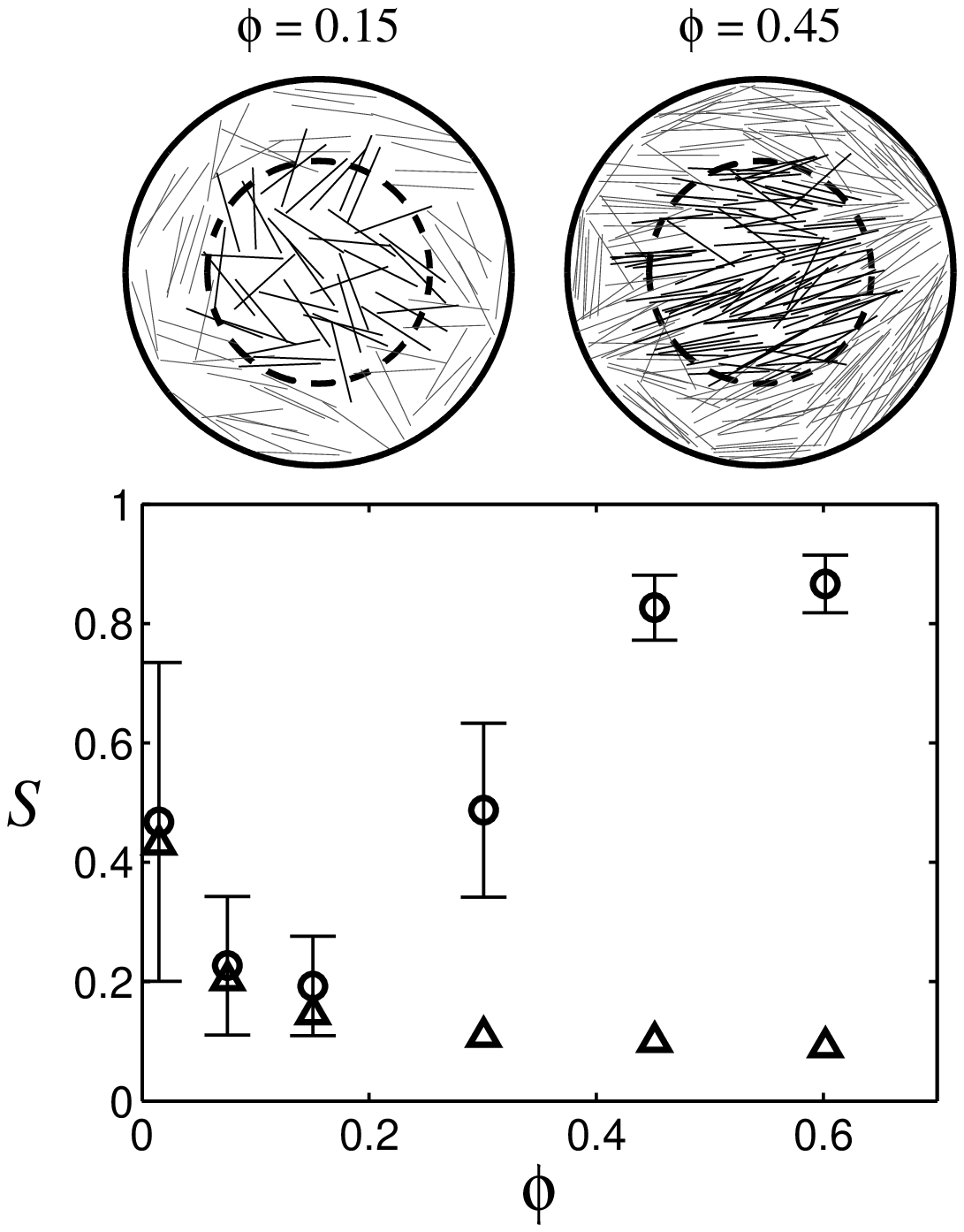}
  \caption{\footnotesize\textsf{Dependence of bulk orientational order on rod density for a single rod aspect ratio $L/D = 40$ in the $R=7\,\rm{cm}$ container. TOP: Representative images, for area fraction $\phi=0.15$ and $0.45$. Dashed lines mark where  $g_{2}(r)\approx 0$, and rods with midpoints within these dashed lines were labeled ``bulk''. BOTTOM: Order parameter, $S$, as a function of $\phi$ for experimental data $( \bigcirc )$. Monte Carlo simulations $( \triangle )$ of non-interacting isotropically oriented rods are included to distinguish rod ordering from the finite-number effect on $S$.}} 
\label{density}
\end{figure}

We first tested the dependence of orientational order on area fraction, $\phi$. Typical experimental results are illustrated in Fig.~\ref{density}, where $L/D=40$ in the $R=7\,\rm{cm}$ container. In the top of Fig.~\ref{density}, the rods in the bulk (highlighted) appear to align as $\phi$ increases. The bottom of Fig.~\ref{density} shows $S$ as a function of $\phi$ as well as $S$ from the MC simulations. For small $\phi$, $S$ is relatively low and coincides with $S$ from simulations, indicating that the experimental system is isotropic. As $\phi$ increases, $S$ rises and deviates from the values derived from simulations, denoting the transition to an ordered nematic state. Conforming to the convention that $S=0$ in an infinite isotropic system, we henceforth subtract the results of the MC simulations from experimental data.

\begin{figure}[!bp]
  \includegraphics[keepaspectratio=true,width=0.29\textwidth]{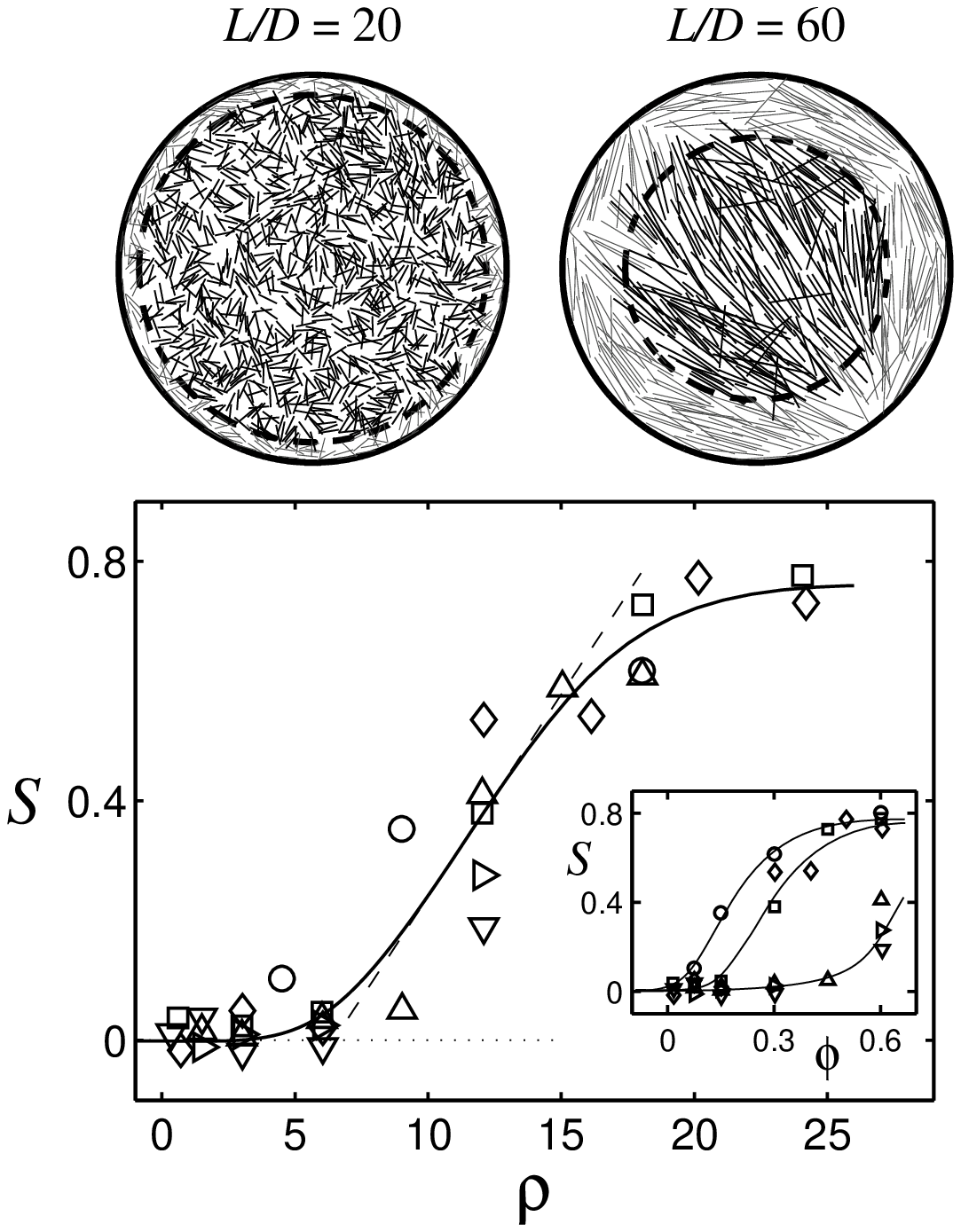} 
  \caption{\footnotesize\textsf{Dependence of bulk orientational order on rod aspect ratio. TOP: Representative images, for $L/D = 20$ and $60$ at $\phi=0.3$ in the $R=15\,\rm{cm}$ container. As in Fig.~\ref{density}, rods within the dashed lines delineate bulk. BOTTOM: Consolidated plot of $S$ vs. scaled area fraction, $\rho=\phi L/D$, for $R=7\,\rm{cm}$ container: $ L/D = 20 (\triangle),\, 40 (\square)$; $R=15\,\rm{cm}$ container: $ L/D = 20 (\rhd),\,  60 (\bigcirc)$; square container: $ L/D = 20 (\bigtriangledown),\,  40 (\Diamond)$. Dashed line is a linear least squares fitting of the data in the transition region. By extrapolating this line to $S=0$, we estimate the transition density between $5$ and $8$. INSET: Same data plotted as $S$ vs. $\phi$. All solid lines are guides for the eye.}}
\label{normalized}
\end{figure}

We then fixed $\phi$ and tested the dependence of orientational order on rod aspect ratio, $L/D$. Experimental results for $L/D = 20$ and $60$ at $\phi=0.6$ in the $R=15\,\rm{cm}$ container are shown in Fig.~\ref{normalized}(top). Though the same mass is present in both containers, the $L/D=60$ rods appear more ordered than the $L/D=20$ rods for this value of $\phi$. The inset in the bottom of Fig.~\ref{normalized} shows $S$ vs $\phi$ for all $L/D$ and containers tested. As $L/D$ increases, the transition to an ordered state shifts to lower $\phi$. 

By plotting $S$ versus the scaled area fraction, $\rho = \phi L/D$, the transition curves for different $L/D$ collapse to a single curve [Fig.~\ref{normalized}(bottom)]. This scaling is consistent with Onsager's mean field result which assumes that the I-N transition stems only from steric interactions between neighboring rods \cite{onsager}. The dashed line is a linear least square fitting of the data in the transition region. By extrapolating this line to $S=0$, we estimate that the critical transition occurs at $\rho=\rho^{\star}$ between $5$ and $8$. For comparison, Onsager's mean field model applied to an infinite $2$D system of rods, predicts an I-N transition at $\rho^{\star}=3\pi/2\approx4.71$ \cite{kayser}, while $2$D computer simulations with periodic boundary conditions estimate $\rho^{\star}$ between $6.5$ and $8$ \cite{frenkel85}. We performed a similar analysis on our data that included both bulk and boundary layer rods. The measured $S$ for the nematic state is systematically lower (due to cancellation by the alignment of the rods along orthogonal walls), but the location of $\rho^{\star}$ remains unchanged.

While the side boundaries do not affect $\rho^{\star}$, the confining walls profoundly influence rod ordering in other ways. Rods near a wall tend to align parallel along it. We assessed this alignment using the wall-rod angular correlation parameter $g_{2}(r)=\langle\,\cos[2(\vartheta_{w}-\vartheta(r))] \,\rangle $, where $\vartheta(r)$ is the angle of a rod located at distance $r$ away from the wall, as described by the shortest line from rod midpoint to wall, and $\vartheta_w$ denotes the vector locally tangent to the wall.

\begin{figure}[!hbp]
  \includegraphics[keepaspectratio=true,width=0.28\textwidth]{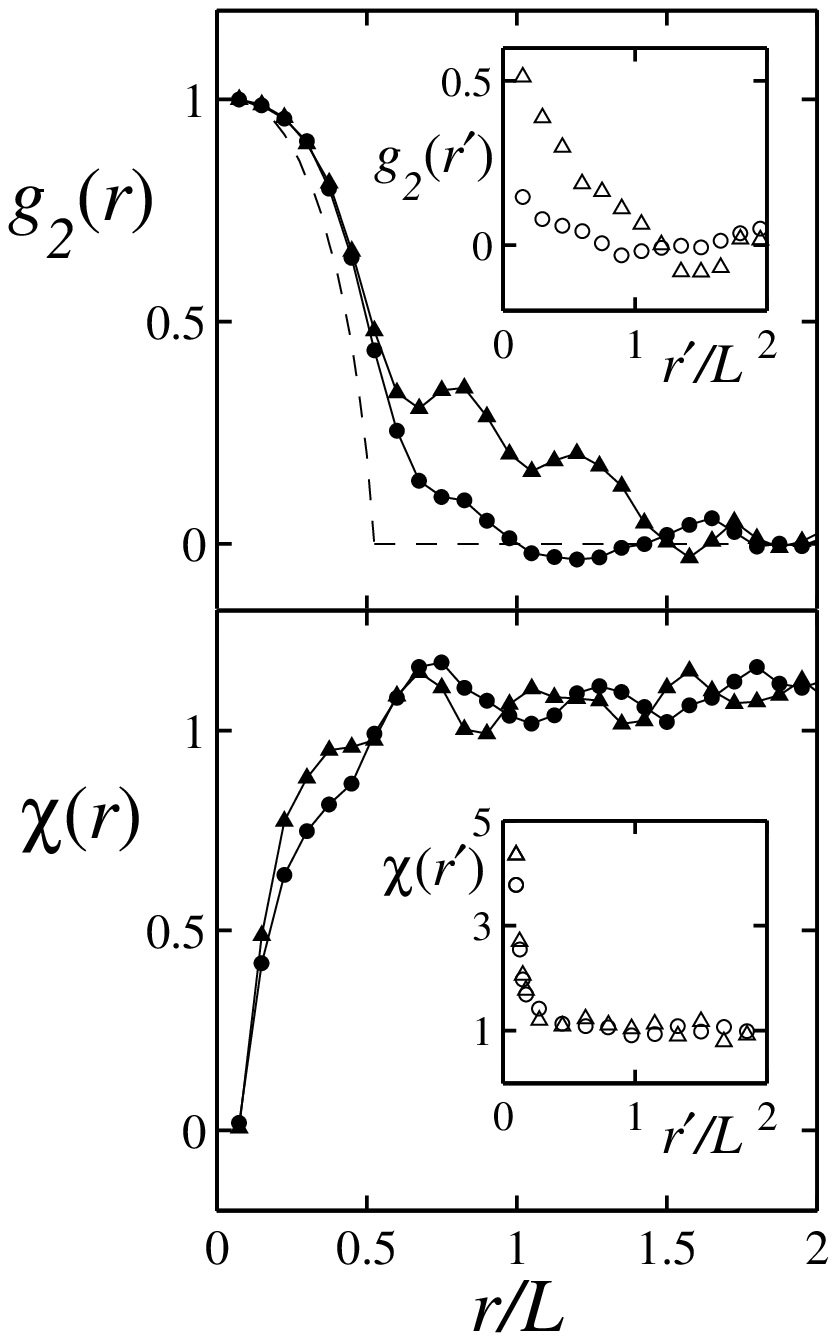} 
  \caption{\footnotesize\textsf{Wall-rod correlation function, $g_{2}(r)$, 
and normalized density profile, $\chi(r)$, 
versus distance $r$ from wall to rod midpoints for 
$L/D=20$ in the $R=7\,\rm{cm}$ container
at $\rho = 1.5$ (circles) and $\rho = 9$ (triangles).
Data was smoothed by a moving average filter with a span of $3$.
Dashed line represents calculated 
$g_{2}(r)$ for a single rod
interacting with the wall via hard core repulsions.
Insets show $g_{2}(r^{\prime})$ and $\chi(r^{\prime})$ versus 
distance  $r^{\prime}$ from wall to rod ends.}} 
\label{wall}
\end{figure}

The top of Fig.~\ref{wall} shows $g_{2}(r)$ for $L/D=20$ in the $R=7\,\rm{cm}$ container at $\rho = 1.5$ (circles) and $\rho = 9$ (triangles). Experiments show an initially high $g_{2}(r)$ that decays as $r$ increases. For comparison, the dashed line represents the MC calculated $g_{2}(r)$ for a single rod interacting only via hard core repulsions with the curved wall. Steric interactions with the wall significantly restrict the possible orientations available for a rod closer than $L/2$ to the wall. Experimental values of $g_{2}(r)$ closely follow the single rod curve for small $r$; however, these values deviate as $r$ increases. Furthermore, as $\rho$ increases, $g_{2}(r)$ diverges earlier and remains elevated for larger $r$. We also measured the number density, $\chi(r)$, normalized with respect to bulk density, of rod midpoints (Fig.~\ref{wall}, bottom). The measured $\chi(r)$ is less than $1$ for $r/L<0.5$, which signifies a depletion of rod midpoints next to the wall. Below $r/L\approx 0.4$, $\chi(r)$ increases and reaches higher values with larger $\rho$. Finally, $\chi(r)$ obtains a local maximum around $r/L\approx 0.6$ before returning to bulk values.
 
Onsager's approximation for excluded volume interactions has been applied to hard spherocylinders in contact with a hard flat wall by Poniewierski \cite{poniewierski}, and results have been confirmed by simulation \cite{dijkstra01}. Although this analysis was performed for a 3D system contacting a planar wall, the experimental features in Fig.~\ref{wall} show striking similarities. The profiles for $g_{2}(r)$ and $\chi(r)$, however, can be difficult to interpret since geometrical constraints tightly couple rod orientation and position for rod midpoints close to the wall. By analyzing distances $r^{\prime}$ based on the closest end of each rod to a wall, rod orientation and position are decoupled, thereby simplifying $g_{2}(r^{\prime})$ and $\chi(r^{\prime})$ profiles \cite{poniewierski}. Fig.~\ref{wall}(top inset) shows a clear $\rho$ dependent increase in $g_{2}(r^{\prime})$ for rod ends near the wall. And, in contrast to the depletion layer seen for rod midpoints, a strong wetting layer becomes apparent, with an approximate four fold increase in $\chi(r^{\prime})$ for small $r^{\prime}$ which decays rapidly to bulk values around $r^{\prime}/L\approx 0.3$ (Fig.~\ref{wall}, bottom inset).

\begin{figure}[!hbp]
  \includegraphics[keepaspectratio=true,width=0.4\textwidth]{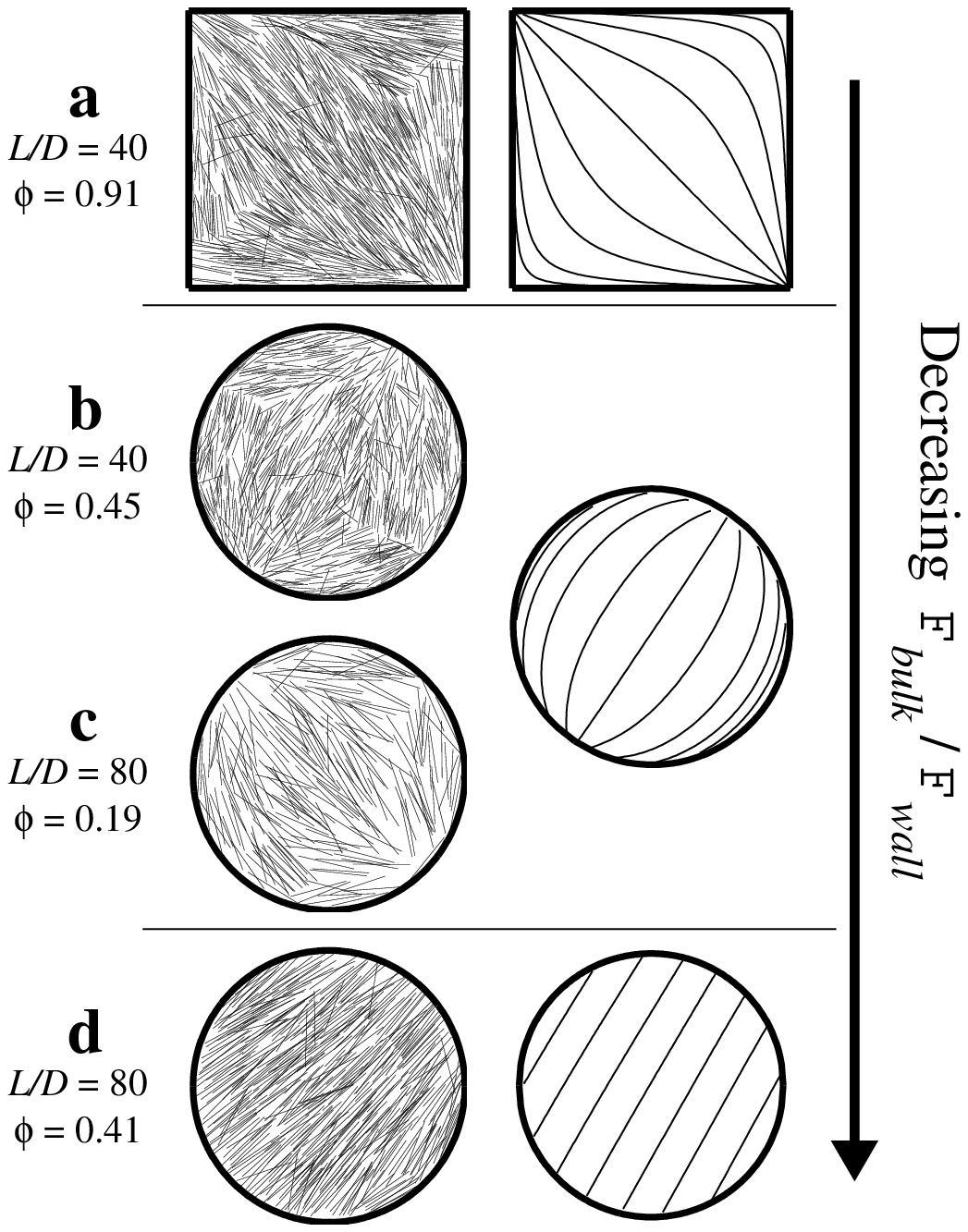}
 \caption{\footnotesize\textsf{Elastic deformations of the nematic
observed for $\rho$ above the I-N transition. Experimental images 
in square and $R=15\,\rm{cm}$ containers (left) 
and results from numeric 
minimization of the elastic free energy functional, $\mathcal{F}$, (right) 
are arranged 
with respect to relative bulk and wall free energies, 
$\mathcal{F}_{bulk}/\mathcal{F}_{wall}$, with 
$\mathcal{F}_{bulk}/\mathcal{F}_{wall}\gg1$, (a),
$\approx1$, (b,c), and
$\ll1$, (d).
This ratio is modified experimentally by changing $L/D$ or $\phi$ (see text). 
A transition can be made from the patterns seen 
in (b,c) to those in (d) by either increasing $L/D$,(b), or 
$\phi$,(c).}}
\label{elastic}
\end{figure}

Although many similarities exist between the hard core interactions of rods with straight versus curved boundaries, wall geometry significantly affects $g_{2}(r)$. Rod alignment with the wall does not penetrate as deeply into the sample center for curved boundaries and, interestingly, the thickness of this layer even regresses at very high $\rho$ (data not shown). This seemingly unusual behavior can be understood by noting that for $\rho \gg \rho^{\star}$ the system must accommodate both the orientation of the nematic director as well as the preferred orientation parallel to the surrounding boundaries. This frustration creates competition between the rods in the bulk with those at the boundary, which produces complex patterns, Fig.~\ref{elastic}(left). 

Even though the local variations in the experimentally observed patterns are often on the order of rod dimensions, we find that these patterns are well approximated by minimizing a continuum free energy functional that describes elastic deformations in ordered liquid crystals \cite{frank,deGennes,priest}. Consider a total free energy composed of bulk and wall contributions, $\mathcal{F} = \mathcal{F}_{bulk} + \mathcal{F}_{wall}$. In the bulk, splay and bend deformations contribute to the distortion energy:
\be 
\mathcal{F}_{bulk}=\frac{1}{2} \int_A  K_{1}(\nabla \cdot \mathbf{n})^2 + K_{3}(\mathbf{n} \times \nabla \times \mathbf{n})^2 ~\mathrm{d}x\,\mathrm{d}y,
\ee
where $\mathbf{n}(x,y) = [\cos \vartheta,\sin \vartheta]$ is the unit vector describing the local nematic director, and $K_{1}$, $K_{3}$ are constants for splay and bend, respectively. Note that contributions from  twist deformations are absent in this $2$D system. The elastic constants relate to the molecular details of the nematic, generally becoming larger with increasing $L/D$ or $\phi$ \cite{priest}. Often, however, the two constants are comparable in magnitude, and here we shall use the simplifying approximation, $K_{1}=K_{3}=K$ \cite{deGennes}. To account for interactions favoring rod orientations locally parallel to the wall, we consider the line integral over the container's perimeter $P$:
\be \label{Fwall}
\mathcal{F}_{wall} = \frac{1}{2} C \int_P \sin^{2}(\vartheta_{w}-\vartheta)~\mathrm{d}s,
\ee
where $C$ is a positive constant with units of line tension, and $\vartheta_w$ describes the vector locally tangent to the wall, $\mathbf{n}_w$. From Eq.~\ref{Fwall}, it becomes clear that $\mathcal{F}_{wall}$ depends not only on anchoring strength, $C$, but also on wall curvature and length.

Fig.~\ref{elastic}(right) shows the results of numeric minimization of the elastic free energy functional $\mathcal{F}$ using a relaxation method with simulated annealing \cite{kirkpatrick}, where the local tangent to each curve represents the orientation of $\mathbf{n}$ at that point. Similar patterns found in experiment (left) and calculations are grouped and arranged according to the relative ratio $\mathcal{F}_{bulk}/\mathcal{F}_{wall}$. As nematic ``stiffness'' increases compared to wall anchoring ({\em increasing} $K/C$), $\mathcal{F}_{bulk}/\mathcal{F}_{wall}$ {\em decreases}, because the system reduces the cost of nematic deformations in the bulk at the expense of misalignment with the wall. 

Experimentally, by holding wall geometry constant and varying $\phi$ and $L/D$, we changed nematic stiffness $K$ and the ratio $K/C$, thus modifying $\mathcal{F}_{bulk}/\mathcal{F}_{wall}$ \cite{deGennes}. Differing combinations of $\phi$ and $L/D$ give rise to equivalent patterns, as seen in Fig.~\ref{elastic}(b,c), corresponding to $\mathcal{F}_{bulk}/\mathcal{F}_{wall}\approx 1$. By raising nematic stiffness though increasing $L/D$ [Fig.~\ref{elastic}(b)] or $\phi$ [Fig.~\ref{elastic}(c)], we transition to the pattern seen in Fig.~\ref{elastic}(d), where $\mathcal{F}_{bulk}/\mathcal{F}_{wall}\ll 1$. At the other extreme, where $\mathcal{F}_{bulk}/\mathcal{F}_{wall}\gg 1$, alignment with the wall is favored at the expense of nematic deformations. By changing wall curvature, we can strengthen wall anchoring and the penalty for wall misalignment, as seen in Fig.~\ref{elastic}(a). For a wide range of parameters, we not only find that the experimental patterns match the theoretical model results, but also that $\mathcal{F}_{bulk}/\mathcal{F}_{wall}$ is sufficient to describe the whole range of nematic deformations witnessed in experiments. However, as may be expected for confined nematics, we have also experimentally observed line and point defects in these phases. \cite{deGennes, dePablo}

To conclude, we have shown that the macroscopic granular rods in this steady state system share many properties with thermally equilibrated lyotropic liquid crystals, such as an I-N transition and steric boundary effects. This system may also be useful for studying defects and other aspects of lyotropic liquid crystals that might be difficult to directly probe experimentally.

We thank D. Lathrop for use of the shaker, and 
J. V. Sullivan for designing experimental components.
This research was supported by the Intramural
Research Program of the NIH (NICHD).


\end{document}